\documentclass[12pt,preprint]{aastex}

\shorttitle{UV escape fraction with LBC}
\shortauthors{Boutsia et al.}

\begin{document}


\title{A low escape fraction of ionizing photons of $L>L^{*}$ Lyman break galaxies at z=3.3}

\author{K. Boutsia$^{1}$, A. Grazian$^{1}$, E. Giallongo$^{1}$, A. Fontana$^{1}$,
  L. Pentericci$^{1}$, M. Castellano$^{1}$, G. Zamorani$^{2}$, M. Mignoli$^{2}$,
  E. Vanzella$^{3}$, F. Fiore$^{1}$, S. J. Lilly$^{4}$,
  S. Gallozzi$^{1}$, V. Testa$^{1}$, D. Paris$^{1}$, P. Santini$^{1}$
\affil{$^{1}$INAF - Osservatorio Astronomico di Roma, Via Frascati 33, I--00040, Monteporzio, Italy }            
\affil{$^{2}$INAF - Osservatorio Astronomico di Bologna, Via Ranzani 1, I--40127, Bologna, Italy}
\affil{$^{3}$INAF - Osservatorio Astronomico di Trieste, Via G.B. Tiepolo 11, I--34131, Trieste, Italy}
\affil{$^{4}$Institute of Astronomy, ETH Z$\ddot{u}$rich, CH-8093, Z$\ddot{u}$rich, Switzerland}
          }


\begin{abstract}
We present an upper limit for the relative escape fraction 
($f_{esc}^{rel}$) of ionizing radiation at z$\sim$3.3 using a sample of 11
Lyman Break Galaxies (LBGs) with deep imaging in the U band obtained with 
the Large Binocular Camera, mounted on the prime focus of the Large Binocular 
Telescope. We selected 11 LBGs with secure redshift in the range $3.27<z<3.35$, 
from 3 independent fields. We stacked the images of our sources 
in the R and U band, which correspond to an effective rest-frame wavelength 
of 1500\AA~ and 900\AA~ respectively, obtaining a limit in the U band image 
of $\ge$30.7(AB)mag at 1$\sigma$. We derive a 1$\sigma$ upper limit 
of $f_{esc}^{rel}\sim 5\%$, which is one of the lowest values found in the 
literature so far at z$\sim$3.3. Assuming that the upper limit for the escape 
fraction that we derived from our sample holds for all galaxies at this redshift, 
the hydrogen ionization rate that we obtain ($\Gamma_{-12}<0.3 s^{-1}$) is not 
enough to keep the IGM ionized and a substantial contribution to the UV background 
by faint AGNs is required. Since our sample is clearly still limited in size, 
larger z$\sim$3 LBG samples, at similar or even greater depths are necessary to 
confirm these results on a more firm statistical basis. 
\end{abstract}



\section{Introduction}
The evolution of the intergalactic medium (IGM) depends primarily on its 
ionization state, which at high redshifts is a function of the
ionizing UV background (UVB) produced by galaxies and AGNs. 
In the early universe, star forming (SF) galaxies are ubiquitous and in principle
can provide the necessary flux to keep the IGM ionized at least up to  
$z\sim6$, when the re-ionization appears to be completed \citep{fan06}. 
However, the poor knowledge of the average escape fraction of ionizing Lyman continuum 
photons from the interstellar medium of each galaxy introduces large
uncertainties to the exact level of contribution in ionizing the IGM. 
The escape fraction of AGN is not strongly variable with redshift 
but the apparent number density of bright QSOs and AGNs is rapidly decreasing
at $z>3$, thus it is assumed that the contribution to the ionizing
flux of the SF galaxies should become dominant at $z>3$ \citep[e.g.][]{fauc08,cowi09}.
In this respect, several attempts have been made to derive the escape
fraction of UV ionizing photons both at low redshifts, from space, and at high 
redshifts ($z=2-4$) from ground based observations. In most cases 
only upper limits on the escape fraction were obtained, giving little 
evidence to support a scenario where enough ionizing photons escape from
galaxies. Actually, early measurements at low and intermediate redshifts, with
$f_{esc}\le$1-3\%, seem to indicate that galaxies are not the major
contributors to the ionizing UV background at $z<2$ leaving AGNs as the 
main ionizing population \citep[e.g][]{gial97,malk03,sian07,cowi10,sian10,brid10}.

At high redshifts direct observations of the Lyman continuum flux for 
galaxies at $z>4$ are difficult because of the sharp increase in the
neutral hydrogen absorption by the IGM. For ground-based surveys, the search
has thus been focused at $z\sim3-4$ which is a compromise between lower detection 
efficiency in the UV and higher transparency of the IGM in comparison to
higher $z$. In this respect, there are two main strategies to face the issue
of escaping ionizing photons from galaxies. One way is to derive direct
constraints using very deep UV imaging and/or spectroscopic observations in 
order to ascertain if even a small fraction of LyC escapes from typical star 
forming galaxies. Such surveys are time consuming, requiring 20-30 hours of 
integration for reasonable UV efficiency, and for this reason involve small 
samples made up of some tens of galaxies. The works by \cite{stei01} and
\cite{shap06} are typical in this respect. Indeed, \cite{stei01} first claimed 
appreciable ionizing flux corresponding to a relative (with respect to
the observed 1500\AA~flux) escape fraction of the order of $>50$\% from a
composite spectrum of the bluest 29 galaxies of their large LBG sample at
$z\sim3.4$. If such an escape fraction were typical of the LBG 
population, this would imply an ionizing UVB at 
least 3 times higher than the one derived from the analysis and modeling 
of the Lyman $\alpha$ absorption by the IGM at the same redshift \citep[e.g][]{bolt05}.
\cite{gial02} obtained the first high S/N long slit spectra of two 
LBGs at $z=3$ and $z=3.3$, selected from Steidel's sample, where no 
significant Lyman continuum emission has been detected. This implies a
$1\sigma$ upper limit to the relative escape fraction of 15\%. Using
individual spectra of 14 LBG galaxies at $z\sim3$ \cite{shap06} found
significant emission only in two galaxies, implying a detected average relative 
escape fraction of 14\% for their sample at a 3$\sigma$ confidence level, 
which is $\sim$4.5 times lower than the value derived by \cite{stei01}. 
The fact that the estimate of the escape fraction value obtained in 
this way has been progressively reducing from 70\% to 15\%, reveals the difficulties 
and the uncertainties involved in such measurements.

Another way to measure escaping ionizing flux is to use shallower UV imaging but of a very large sample of
galaxies. In this case, instead of measuring the average/typical escape fraction of the population,
the effort is concentrated in finding possible small fractions of special
types of galaxies which could show a large percentage ($>$50\%) of escaping
ionizing flux. The main limitations of this method are the
absence of extensive spectroscopic confirmation of the redshifts in the
sample or, whenever spectral information is available, the large redshift range of the sources, which
requires extended simulations to correctly account for the variances of IGM
absorption, and the lower S/N level of the narrow band (NB) images.
Typical in this direction is the work by \cite{iwat09}, who have used a UV narrow band filter technique
and found ionizing radiation, at a 2$\sigma$ level, in 7 out of 73 LBGs at 
$z>3$ LBGs and 10 out of 125 Ly$\alpha$ emitters at $z>3$ out of a sample 
of 198 $z>3$ galaxies in the SSA22 field. The relative escape fraction derived 
for these 7 detected LBGs reaches 80\%. It is interesting to note that they 
report null emission for the object SSA22a-D3, for which \cite{shap06} claimed an 
ionizing flux detection. Moreover in several cases \citep[including the second object
detected by][]{shap06}, an offset is present between the position
of the object in the UV narrow band filter and in the R band detection
image, suggesting possible contamination by interlopers.
\cite{vanz10a} evaluated the probable cause of such offsets, showing
that there is a $>50$\% probability for contamination, by lower $z$ objects, 
in at least 30\% of the galaxies showing clear Lyman continuum detection. 
Recently, \cite{vanz10b} using ultra-deep VIMOS
intermediate U band and deep FORS1 narrow band imaging of 102 galaxies in the
GOODS-South at $z\sim$3.7, obtained an $f_{esc}$ upper limit of 5\%-20\% at 
a 3$\sigma$ level, depending on the assumed extinction 
curve and E(B-V) value, as well as the luminosity and redshift of the sources 
adopted in the analysis. 

These two approaches should be considered as complementary, since they give us
information on different aspects of the problem: the typical escape fraction
from the majority of galaxies, the former, and the existence of rare galaxies
with higher escape fractions, the latter. Both methods are necessary to 
measure with greater accuracy the global amount of ionizing flux in the universe.
In the present paper we use deep U and R band imaging obtained by the UV
optimized Large Binocular Camera \citep{gial08} in two Steidel fields and in 
a region of the COSMOS field to derive stringent limits on the ionizing 
escape fraction of 11 LBGs at $z\sim 3.3$. The uniqueness of our data is 
represented by the simultaneous availability of very deep UV images, 
obtained with one of the most efficient ground based large field UV imager at an 8m 
class telescope, in fields where spectroscopically confirmed galaxies in a 
very narrow redshift range around $z\sim 3.3$, are already available. 
This sample, although it is limited in size, gives us the possibility to 
focus on a very small redshift range, reducing appreciably the effect of 
the variation with redshift of the IGM opacity and allowing a more direct 
measure of the relative ionizing escape fraction from LBGs at $z\sim 3.3$.

\section{Data}
The targets were selected from three fields where deep UV images were 
obtained with the Large Binocular Camera (LBC) at the Large Binocular 
Telescope \citep[LBT,][]{hill10}. The LBC has two channels, one optimized 
for the UV and blue bands and one optimized for imaging in the red bands, that was used to obtain the R band
images. The detectors have four 4K$\times$2K chips with a pixel scale of 0.23
$arcsec/pixel$, providing an unvignetted field of view of about $23\times 23 arcmin^{2}$. 
The standard LBC pipeline \citep{gial08} has been used for data
reduction. In particular, after creating the bias-subtracted
images, we have applied standard flat-fielding using sky flats obtained during 
twilight. The sky background was subtracted using SExtractor \citep{bert96} 
with a background mesh of 64$\times$64 pixels and adopting a median filter of 3$\times$3. 
The astrometric solution was computed using the software AstromC developed by \cite{rado04} and the
USNO-A2 \citep{mone98} as reference catalog. After applying the
astrometric correction, we have coadded the various images to create deep mosaics for each field. 
We did not include a color term in the photometric calibration and the
accuracy of the zeropoint is 0.03-0.05 for the U filter. 
Part of these images have already been used to derive the
number counts in the U band by \cite{graz09} providing good agreement with
previous results. 

For each scientific image the LBC pipeline computes a corresponding RMS map
directly from the raw science frames. We obtained the absolute RMS maps 
for each pointing by computing the RMS in each individual image 
(using Poisson statistics and the instrumental gain as $\sigma_{i}=\sqrt{\frac{N_{ADU}}{gain}}$) and 
self-consistently propagating this RMS over the whole data reduction
process. The delivered RMS maps were found to be consistent with the absolute RMS
expected for each image and with the median RMS derived from 5$\times$5 pixel
boxes in random positions of the sky (see \cite{gawi06} for details).

Two LBC fields come from Steidel's sample and include the bright quasars 
Q0933+28 ($\alpha$=09:33:31, $\delta$=+28:44:42) and Q1625+26 
($\alpha$=16:25:30, $\delta$=+26:52:31) \citep{stei03,redd08}. A major 
advantage that led to the selection of these areas is the existence of 
several spectroscopic redshifts, obtained by \cite{stei03} in the redshift 
range $2<z<3.5$. The LBC images were obtained in the period 2007-2009 in 
the U band with total exposure times of 12h and 8h, respectively, and in the R 
band with total exposure times of 2.8h and 3.4h. The AB magnitude limits
reached at 1$\sigma$ in the U band are 28.8 and 28.6 and in the R band 25.5 and 25.9 at 10$\sigma$, 
respectively, using an aperture of 2$\times$FWHM. The seeing in the R band is 1.0$\arcsec$ for Q0933+28
and 0.8$\arcsec$ for Q1625+26. For both fields, the average seeing in the U
band is 1.1$\arcsec$.

The third LBC field is within the COSMOS field \citep{scov07,lill07} 
where extensive multiband and spectroscopic dataset is continuously 
improving. The LBC observations of the COSMOS field started in February 
2007 to add deep UGRIZ imaging in the central 1 deg$^2$ and are currently 
on-going. The exposure time per pointing in the U band is $\sim$6 hours, 
for a total of three partially overlapping LBC pointings, reaching 11 hours at
the deepest region. Actually, the whole area covered by deep $UGRIZ$ data in 
COSMOS is $\sim$700 sq. arcmin. and reaches a 1$\sigma$ AB magnitude 
limit of U=28.7, with a typical seeing of 1.0$\arcsec$. The R band image has an
average seeing of 0.9$\arcsec$ and it reaches an AB magnitude limit of R=25.9 at
10$\sigma$, for two pointings. The third pointing was obtained
during commissioning and it has a total exposure time of 0.5 hours, an average
seeing of 0.9$\arcsec$ and an AB magnitude limit of R=23.8. The datasets 
in all three fields are composed by different mosaics, obtained in different
years and in a variety of seeing conditions and exposure times. Although we 
have average seeing values estimated in each band, these values do not remain 
stable throughout the whole final mosaic. Actually, the seeing in the R band 
ranges from 0.8$\arcsec$ to 1.0$\arcsec$ and in the U band from 1.0$\arcsec$ to 1.2$\arcsec$,
especially in the COSMOS field where we cover the largest area and we 
have 3 different pointings with different final exposure times. 

\section{Analysis}
It is important to note that there are several definitions for the 
ionizing escape fraction which depend on the non-ionizing flux considered as a 
reference. The absolute escape fraction is the ratio between the escaping
Lyman continuum flux and the one intrinsically produced by stars in the galaxy \citep{leit95}. 
In practice this value is difficult to determine because we should know 
the intrinsic Lyman continuum flux from an accurate fit to the overall
spectral energy distribution of the galaxy. For this reason, observationally, the
ionizing escape fraction is usually related to the rest frame 1500\AA~ flux
and denoted as $f_{esc}$. This reference flux is however attenuated by dust
and a correction is needed if we want to derive the intrinsic Lyman 
continuum. If the amount of dust attenuation is not known in the analyzed 
galaxy sample, a relative escape fraction $f_{esc}^{rel}$ can be introduced as
the fraction of escaping Lyman continuum photons divided by the fraction of 
escaping photons at 1500\AA~\citep{stei01}. The relation between the two
quantities is $f_{esc} =f_{esc}^{rel} \times10^{-0.4A_{1500}}$ where $A_{1500}$ 
is the dust absorption at 1500\AA~ \citep{inou06,vanz10b}. Since LBGs at z=3 show an average
$A_{1500}$=0.6 \citep{vanz10b} then $f_{esc}^{rel}$ should be  typically 2 times larger than $f_{esc}$.
Finally, to derive the Lyman continuum ionizing fraction from the observed
fluxes at 900\AA~ and 1500\AA~, we need to estimate the contribution to the
absorption by the intervening IGM and the average
intrinsic UV spectral shape of the galaxy populations. 
In our analysis we estimated the relative escape fraction using the following equation:

\begin{equation}
{f}_{esc}^{rel}=\frac{({L}_{1500}/{L}_{900})_{int}}{({f}_{1500}/{f}_{900})_{obs}}exp(\tau^{IGM}_{900})
\end{equation}

where $({L}_{1500}/{L}_{900})_{int}$ is the average intrinsic ratio of
non-ionizing to ionizing specific intensities as derived from spectral   
synthesis models \citep[e.g.][]{bruz03}, $({f}_{1500}/{f}_{900})_{obs}$ is the observed flux ratio
estimated from our U-R color and $exp(\tau^{IGM}_{900})$ is the inverse of the
average IGM transmission at $z\sim3.3$. For comparison with previous estimates of ${f}_{esc}^{rel}$ in the 
literature, we adopt $({L}_{1500}/{L}_{900})_{int}$=3, but this value
is model dependent and will be discussed below. 
A major issue in the derivation of the relative escape fraction is the estimate of the 
average attenuation produced by the IGM along the line of sight, $exp(\tau^{IGM}_{900})$.
This value is very sensitive to the actual redshift distribution of the
sources and to the wavelength bandpass of the filters used. The broader the
filter, the more the luminosity of the continuum at $\lambda \leq$ 912\AA~ will
be diluted, resulting to a less constraining value for the escape
fraction.

The average IGM transmission $exp(-\tau^{IGM}_{900})$ for our
galaxy sample has been derived following the work of \cite{proc09}, where they analyzed the
average rest-frame spectra of z$>$3 SDSS quasars. They measured the IGM
absorption from the drop of the flux at $\lambda\leq$912\AA, rest frame, in a
redshift range 3.6$<z<$4.3. To reproduce the IGM attenuation at $z\sim$3.3, instead of 
using the Lyman absorption statistical distribution in column density and redshift, 
we decided to adopt the same empirical fit described by the authors. 
More specifically, since the lowest redshift they consider is
3.6, we extrapolated the IGM LyC absorption to the average redshift 
of our sample (z=3.3) using the formula (Eq.6) given by \cite{proc09}, 
and adding the contribution of the Lyman series absorption using the
equations described by \cite{fan06}. The attenuation 
of photons that were emitted at an observed wavelength,
$\lambda_{obs}$, at redshift $z_{em}$=3.3, is shown in Fig.~1 (solid black line).

The LBC UV broad band transmission covers the wavelength range 
from 3200-3900\AA, with a small tail up to 4000\AA~(considering an average
airmass for our dataset of 1.2 and including instrument and CCD response). 
This means that Lyman continuum emission by LBGs at $\lambda<$912\AA~starts just short-wards of
the red edge of the U filter for sources with $z\simeq 3.3$, leaving the LBGs 
non ionizing emission outside the filter at $\lambda_{obs} >$ 4000\AA~. 
The effective rest-frame wavelength for the Prochaska IGM attenuation, 
convolved with the LBC U band filter and considering z=3.3, is 860\AA~, which is
reasonably close to 900\AA~. Thus, in order to avoid contamination from
continuum emission red-wards of the Lyman limit and to probe a wavelength close to 
900\AA~, without introducing excessive dilution of the observed emission due 
to IGM attenuation, we decide to use only LBGs with spectroscopic redshifts in the range $3.27<z<3.35$.
We consider this as the best trade-off in order to obtain tight constraints for the escape fraction.

\begin{figure}
\includegraphics[width=10cm,angle=0]{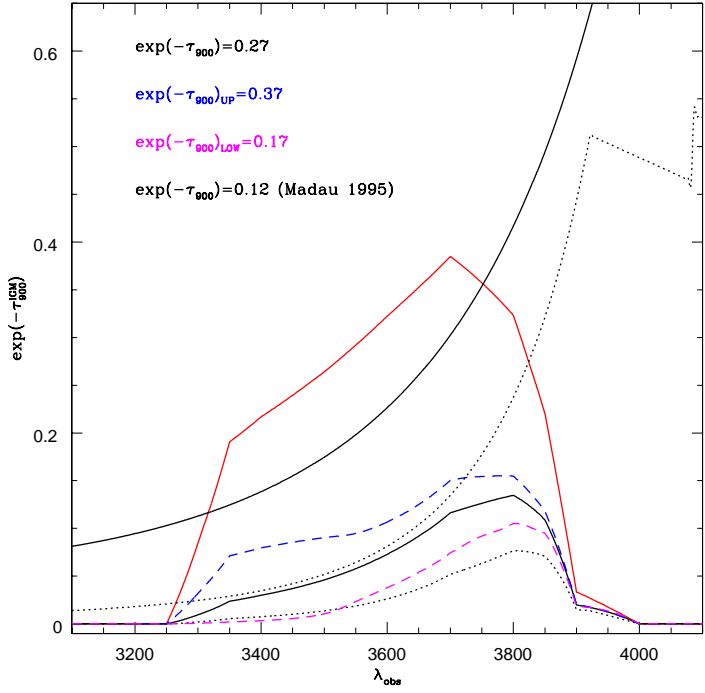}
\caption{The IGM transmission is calculated using the \cite{proc09} value
  extrapolated at z=3.3 (black solid line) along with the total transmission
  curve for our U band (red solid line). The lower black dotted line
  corresponds to the Madau IGM transmission and it is shown for comparison
  only. The solid and dotted lines under the filter transmission are the resulting
$exp(-\tau^{IGM}_{900})$ curves after the convolution with the filter response
  using the Prochaska and Madau IGM transmission curves respectively. 
The color-coded lines correspond to the upper and lower limit of the IGM attenuation showing the uncertainty in
our measurement as given by the work of \cite{proc09} (see parag.~4 for
  details). The filter transmission and the resulting $exp(-\tau^{IGM}_{900})$
  curves after the convolution have all been multiplied by a factor of 2 to facilitate the view. At
$\lambda\approx$ 3700 the real filter transmission value is 0.19.}
\label{fig:exptau}
\end{figure}

We have a total of 12 LBGs in the tight redshift range ($3.27<z<3.35$) in all
of our fields: 4 sources in Q0933, 1 in Q1623 and 7 in COSMOS. In the COSMOS 
field one of the sources ($\alpha$=10:01:19.29, $\delta$=+02:04:20.2) showed 
significant emission in the U band ($>3\sigma$), which was slightly offset 
from the image center in the R band. We checked the high resolution ACS 
image finding that the morphology of the source appears more consistent with 
two distinct objects, which are too close to be resolved in ground based 
observations. Careful inspection of the spectrum also supports such a scenario, 
where the main spectral features clearly appear only in the UV dropout
candidate. In the LBC images there is an offset of $\sim0.6\arcsec$, between 
the emission in the R and U band that corresponds to $\sim$6kpc. For all these 
reasons, we decided not to include this source in our sample, leaving us with
a total of 11 LBGs in the 3 fields. According to \cite{vanz10a}, assuming the
number counts for U$<$28.5 with a seeing of 1$\arcsec$, there is more than
80\% probability that 10\% of the sources are affected by contamination from
lower redshift sources in the line of sight. Thus, the fact that 1 out of our
12 LBGs seems to be contaminated is within the statistical limits.     

\begin{figure}
\includegraphics[width=9cm,angle=0]{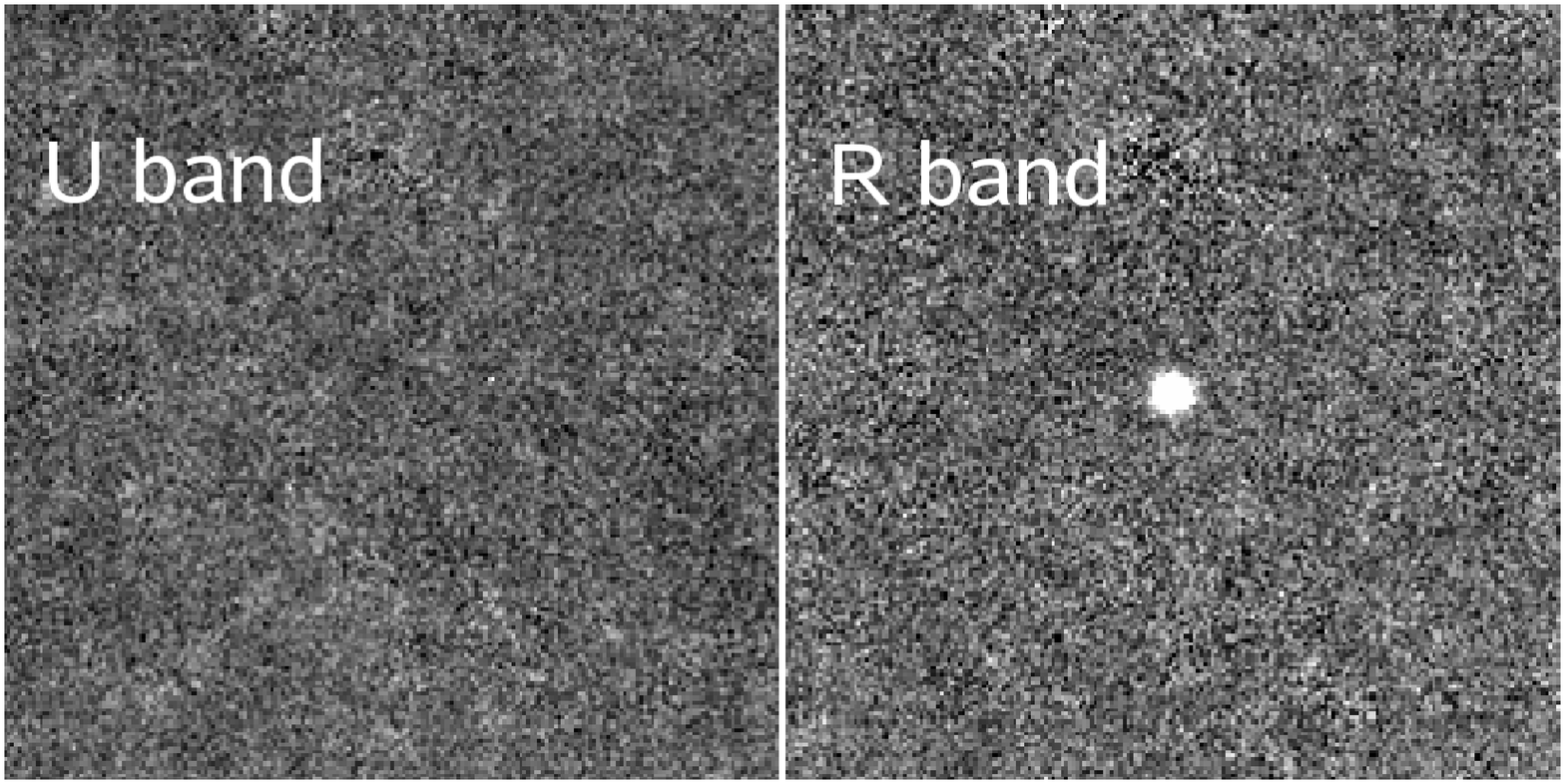}
\caption[The stacking of eleven LBGs]{The stacking result for the 11 LBGs of
our sample. The size of each thumbnail is $45\arcsec\times45\arcsec$}
\label{fig:stackURall}
\end{figure}

To compute a reliable upper limit to the Lyman continuum escape 
fraction we created thumbnail images of 45$\arcsec\times45\arcsec$ in the U and R bands 
around each of the selected sources. The size of the thumbnail is large
enough to obtain a reliable estimate for the background, which is subtracted
after masking the surrounding objects. We then sum all thumbnails in each band 
creating a stacked image, where the flux corresponds to the weighted mean value 
of the fluxes of the individual images in each band. 
In Fig.~\ref{fig:stackURall} we show the result of this procedure. 
We obtained aperture photometry around the source in the R band (${f}_{1500}$) 
where the source is visible, extracting the center of the stacked profile. 
This position is then used to measure the expected aperture flux 
in the U band image (${f}_{900}$). Fig.~\ref{fig:stackURall} shows no
significant flux in the U band stacking. In order to establish the optimum apertures for the photometry 
we created a stacking of stars. The selected stars are near our
candidates in the 3 fields, and their flux has been normalized in order to get
the same magnitude distribution of the LBG stacking. This is necessary since
the magnitudes of the LBGs used vary significantly and the characteristics
of the brightest sources will be more prominent in the final stack. The average measured
stellar FWHM is 0.9$\arcsec$ in the R and 1.1$\arcsec$ in the U band. Using the 
stellar PSF, we have created a kernel that matches the PSF in the R band to
the one in the U band. Once calculated, such Gaussian filter has been applied 
to the R band image of the LBG stacking, in order to match the R band
morphology to the expected U band size. The finally adopted aperture diameters 
correspond to 1.5$\times$FWHM in U band, namely, 7.3 pixels. The magnitude
measured in the smoothed R band is 24.35 mag (AB). The 1$\sigma$ rms
background error in the U band has been used as the flux upper limit and it 
is equivalent to a lower limit in magnitude of $U\ge30.7(AB)$ at $1\sigma$ 
confidence level for the LBG stacking.

\section{Results and Discussion}

Using the 11 sources with spectroscopic redshift z$\sim$3.3 
we obtain $({f}_{1500}/{f}_{900})_{obs}\geq$224. Considering the 
average IGM attenuation extrapolated at this redshift from the Prochaska 
value ($exp(-\tau^{IGM}_{900})$=0.27), we derive ${f}_{esc}^{rel}\le$0.05 (5\%) 
at 1$\sigma$.
Compared to the values in the literature so far, 
this is one of the most constraining results for this redshift range. 
An additional advantage of our sample is that it covers 3 independent areas 
in the sky and includes spectroscopically confirmed LBGs that have been selected through 
different methods. More specifically, the LBGs in the Q0933 field are U 
drop-out sources, the one in the Q1623 field is a BX candidate \citep{redd08} 
and out of the 6 sources in COSMOS, 4 have been selected as both U drop-out and gzK 
candidates, while 2 are only U drop-out candidates.

For a more detailed approach, we used the same technique to estimate 
the escape fraction also for individual sources. The resulting values 
are presented in Table \ref{tab:ind}. The average magnitude of the 
sources is $\sim$24.4 in the R band and the values of the upper limits 
of the relative escape fraction for individual sources range from $<$6\% to $<$36\%, 
with an average value of $<$21\%. To exclude the possibility that our result is 
biased by the wide range in the R magnitude we created a subset of our sample, 
where we used for the stacked image only the sources with $23.5<R<24.7$ (6 sources). 
In this case we obtain a relative escape fraction of $<$6\%.

\begin{table}
\caption{Summary of the ${f}_{esc}^{rel}$ values for the individual LBGs.}
\label{tab:ind}
\centering
\begin{tabular}{ccccccc}
\hline\hline
ID &RA&DEC&$z$&Rmag&Umag&${f}_{esc}^{rel}$\\ 
   &  &   &   &$\pm$0.07& 1$\sigma$ (u.l)& 1$\sigma$ (u.l) \\
\hline
 3400 & 143.35424& +28.80694& 3.27 & 24.88& 29.75 & 0.203\\ 
12646 & 143.32868& +28.71913& 3.33 & 25.13& 29.61 & 0.329\\ 
 8556 & 143.38236& +28.75308& 3.33 & 24.91& 29.66 & 0.258\\ 
10849 & 143.36004& +28.73414& 3.35 & 25.59& 29.66 & 0.357\\ 
17175 & 246.46910& +26.89244& 3.34 & 24.69& 29.05 & 0.235\\ 
74113 & 149.88620& +2.276064& 3.33 & 23.56& 29.66 & 0.062\\ 
50989 & 149.83421& +2.416729& 3.31 & 24.61& 29.25 & 0.223\\ 
32388 & 149.77887& +2.229502& 3.30 & 24.63& 29.60 & 0.163\\ 
51227 & 149.89208& +2.414816& 3.28 & 24.45& 29.28 & 0.202\\ 
 1723 & 150.44702& +2.347633& 3.30 & 23.63& 28.23 & 0.234\\ 
13903 & 150.41495& +2.158999& 3.29 & 22.59& 28.79 & 0.056\\ 
\hline
stack &   --       &    --  & 3.3 & 24.85& 30.73 &  0.050\\    
\hline
\end{tabular}
\end{table} 

As already stated in the introduction, there are two main uncertainties
involved in the measurement of an accurate escape fraction, concerning
the transmission of the IGM and the intrinsic $L_{1500}/L_{900}$ ratio. 
The uncertainty associated to the IGM is computed 
following the work of \cite{proc09}. Using the 1$\sigma$ limits they state for the IGM
attenuation, we compute upper and lower limits for the value of $exp(\tau^{IGM}_{900})$. 
This way, we obtain $exp(-\tau^{IGM}_{900})_{upper}$=0.37 
and $exp(-\tau^{IGM}_{900})_{lower}$=0.17, from which we derive 
$({f}_{esc}^{rel})_{lower}\leq$0.04 (4\%) and $({f}_{esc}^{rel})_{upper}\leq$0.08
(8\%) respectively. For comparison with early results in the literature we
derive also the relative escape fraction using the Madau model. As seen in Fig.1 the
$exp(-\tau^{IGM}_{900})_{Madau}$ is 0.12 and we obtain ${f}_{esc}^{rel}\leq$0.11 (11\%).
Since Madau's first analysis, the statistics of IGM absorption have changed
appreciably \citep[e.g][]{fan06,meik06,inou08} and 
Madau's model represents now a lower limit to $exp(-\tau^{IGM}_{900})$ \citep{inou08}, 
making the ${f}_{esc}^{rel}$ value based on this model an upper bound.
In fact, considering all uncertainties we find that the highest upper limit for the 
escape fraction in no case exceeds 11\%, that corresponds to twice our fiducial
value.

Another source of uncertainty in estimating the relative escape fraction is
the assumed value for the parameter $L_{1500}/L_{900}$. So far, the most widely 
used value is $L_{1500}/L_{900}$=3 \citep[e.g.][]{stei01,inou05,shap06} and we 
have adopted it in order to facilitate the comparison of our results with previous 
works in the literature. Such value is derived considering typical models 
of UV SED for star-forming galaxies with constant star formation rate (SFR). 
However, this ratio of the intrinsic luminosities depends also on the age of the
galaxy and on its star formation history (SFH). Even for constant SFH,
\cite{inou05} find that the intrinsic ratio depends on the time passed since 
the onset of star formation and ranges from 1.5 to 5.5, approaching the upper
edge for older ages. More specifically \cite{meik09} has found a 
$f_{\nu}$(1500\AA~)/f$_{\nu}(912$\AA~) that ranges from 2.7 considering 1Myr
starbursts to 6.2-6.7 for 100-300Myr that are the typical ages of LBGs.
If we consider an intrinsic luminosity ratio of 7, that is the value
recently used by \cite{vanz10b}, our estimated escape fraction would increase
by a factor of 2.3 (${f}_{esc}^{rel}\leq$11.5\%).

Nevertheless, assessing the role of LBGs in producing the ionizing
UVB does not depend on the assumed value for the parameter
$L_{1500}/L_{900}$. Following the work by \cite{inou06}, where they define the
Lyman continuum to UV escape flux density ratio, the luminosity density
${\rho}_{900,esc}$ corresponds to:
\begin{equation}
{\rho}_{900,esc} = {\rho}_{1500} \frac{L_{900}}{L_{1500}} f_{esc}^{rel} = {\rho}_{1500} (\frac{f_{900}}{f_{1500}})_{obs} exp(\tau^{IGM}_{900})
\end{equation}
After converting the luminosity density uncorrected for dust extinction derived by \cite{redd09} at
1700\AA~ to 1500\AA~, we obtain 
$\rho_{1500}=(3.61\pm0.24)\times10^{26}erg s^{- 1} Hz^{- 1} Mpc^{- 3}$, which
corresponds to a $\rho_{900,esc} = 5.98\times10^{24} erg s^{- 1} Hz^{- 1} Mpc^{- 3}$.
We note that this luminosity density has been
obtained using the slope of the luminosity function (LF) given by \cite{redd09}, $\alpha=-1.73\pm0.13$, and 
integrating this LF down to their faint magnitude limit ($M_{AB}(1500)$=-17.6 corresponding to R$\sim$28), 
assuming that the ${f}_{esc}$ does not evolve with luminosity. If we consider sources with R$\le$25.5, 
that is the magnitude limit of our sample ($M_{AB}(1500)$=-20.2), we obtain $\rho_{1500}=1.27\times10^{26}erg
s^{- 1} Hz^{- 1} Mpc^{- 3}$, from which we derive $\rho_{900,esc} =2.1\times10^{24} erg s^{- 1} Hz^{- 1} Mpc^{- 3}$.  
At $z\sim3$ \cite{bolt05}, using high resolution simulations to model the 
Ly$\alpha$ forest opacity, estimate the hydrogen ionization rate of the IGM 
as $\Gamma_{-12}=0.86^{+0.34}_{-0.26}s^{-1}$ (ionization rate in $10^{-12}$ $s^{-1}
atom^{-1}$), which corresponds to a luminosity
density of $\rho_{900}=16.5\times10^{24} erg s^{- 1} Hz^{- 1} Mpc^{- 3}$. 
The values presented by \cite{bolt05} are in good agreement with more recent
works \citep[e.g.][]{agli09}. Our estimated contribution, derived from $M_{AB}(1500)<$-20
sources, is significantly lower than the theoretical value.
Even considering the contribution of fainter galaxies, the estimated emissivity does not 
increase enough in order to supply the required ionizing radiation. Due to our tight 
constraint on the escape fraction, we derive an upper limit to the hydrogen ionization 
rate $\Gamma_{-12}<0.3 s^{-1}$, using as an average spectral index
$\alpha_{UV}$=1.8 \citep{fauc08} and $M_{AB}(1500)\le$-17.6. This is lower than the values 
reported by \cite{bolt05} and only including their estimated AGN contribution
at z$\sim$3 ($\Gamma_{-12}=0.4 s^{-1}$), the ionization rate is barely
consistent with the theoretical expectations from the Ly$\alpha$ forest
analysis. 

Here we should discuss the possibility of a bimodal distribution of the 
LyC escape fraction, where a small fraction of the galaxy population at a 
given redshift can be the major contributor to the cosmological ionizing flux. 
At present, our sample of 11 LBGs is still statistically modest and
cannot exclude in a conclusive way the existence of a limited sub-group of
galaxies with significant escape fraction. Moreover, the search for such rare 
strong LyC emitters at high redshifts is not as straightforward as at lower redshifts. 
Indeed, \cite{cowi09} found only 1 galaxy out of 600 presenting a significant 
fraction of escaping ionizing flux at $z\sim1$. At higher redshifts, \cite{iwat09} 
claims that up to $\sim$10\% of galaxies at $z\sim3.1$ show escape fractions at 
the level of $\sim80$\%. However, an offset ($1\arcsec-2\arcsec$) between the image 
in the R band, sampling the non-ionizing flux at $\sim$1500\AA~and the UV images, 
sampling the LyC flux, has been noted by the same authors. Since the 
escape fraction derived by the UV-R color in the detected LyC emitters is at 
a the level of $\sim80$\%, this would imply an unrealistic fraction, $>100$\%, 
from the limited spatial region where the claimed LyC emission is observed. 
Through statistical analysis, this number has been questioned by \cite{vanz10a}, 
claiming that at least 30\% of those galaxies have more than 50\% probability of 
being low redshift interlopers. On the other hand, \cite{vanz10b} have found only 1 
galaxy out of 102 LBGs at $3.4<z<4.5$ that shows strong LyC emission. In any case, 
assuming a population of strong LyC emitters were present at z=3, this could be at 
most of the order of 1-7\% for the overall population. Adopting this 
contamination fraction as a reference value, the contribution of this subsample 
to the UV emissivity translates to an increase in the ionization rate of the order of 
$\Gamma_{-12}$=0.1-0.3$s^{-1}$, to be added to our original upper limit 
($\Gamma_{-12}<0.3 s^{-1}$). Even in this case, the ionization rate is still 
not sufficient to keep the IGM ionized, making necessary an appreciable 
contribution by some other population, possibly AGN or a large number of very 
faint star forming galaxies with increasingly larger escape fractions.

Such a scenario, where the AGN contribution is still important at
$z>3$ is also supported by the work of \cite{sian08}, who derive an
ionization rate of $\Gamma_{-12}=0.48s^{-1}$ for QSOs at $z\sim3.2$,
suggesting that AGN and LBGs are emitting comparable fractions of ionizing flux to the
intergalactic medium. This is in agreement with recent results by \cite{glik11}, who find that 
the AGN account for at least half the ionizing radiation needed at these
redshifts (60$\pm$40\%). Moreover, our observed limit for the $f_{esc}^{rel}$, 
is closer to the values reported by large surveys at
lower redshifts ($1<z<2$) suggesting a milder evolution with redshift. 
Deeper UV and narrow band observations of larger statistical samples of LBGs 
at z$\sim$3 will allow us to clarify this issue. 


\acknowledgments
K.B. would like to thank N. Reddy for providing redshifts for the sources in the Steidel fields.  
Observations have been carried out using the Large Binocular Telescope
at Mt. Graham, Arizona. The LBT is an international collaboration among
institutions in the United States, Italy and Germany. LBT Corporation
partners are: The University of Arizona on behalf of the Arizona
university system; Istituto Nazionale di Astrofisica, Italy; LBT
Beteiligungsgesellschaft, Germany, representing the Max-Planck
Society, the Astrophysical Institute Potsdam, and Heidelberg
University; The Ohio State University, and The Research Corporation,
on behalf of The University of Notre Dame, University of Minnesota and
University of Virginia.

\end{document}